# Structural and optical properties of self-catalytic GaAs:Mn nanowires grown by molecular beam epitaxy on silicon substrates

Katarzyna Gas,[a] Janusz Sadowski*[b,a], Takeshi Kasama[c], Aloyzas Siusys[a], Wojciech Zaleszczyk[a], Tomasz Wojciechowski[a], Jean-François Morhange[d‡], Abdulmenaf Altintaş[e], H. Q. Xu[e,f], Wojciech Szuszkiewicz[a]

[a] Institute of Physics, Polish Academy of Sciences, al. Lotników 32/46, PL-02-668 Warszawa, Poland

[b] MAX-IV laboratory, Lund University, Box 118, SE-221 00 Lund, Sweden

[c] Center for Electron Nanoscopy, Technical University of Denmark, DK-2800 Kongens Lyngby, Denmark,

[d] Institut des Nanosciences de Paris, UMR 7588, UPMC, 4 pl. Jussieu, 75252 Paris cedex 05, France,

[e] Division of Solid State Physics and the Nanometer Structure Consortium, Lund University, Box 118, SE-221 00 Lund, Sweden,

[f] Key Laboratory for the Physics and Chemistry of Nanodevices and Department of Electronics, Peking University, Beijing 100871, China

Mn-doped GaAs nanowires were grown in the self-catalytic growth mode on oxidized Si(100) surface by molecular beam epitaxy and characterized by scanning and transmission electron microscopy, Raman scattering, photoluminescence, cathodoluminescence, and electron transport measurements. The transmission electron microscopy studies evidenced the substantial accumulation of Mn inside the catalyzing Ga droplets at the top of the nanowires. Optical and transport measurements revealed that the limit of Mn content for self-catalysed growth of GaAs nanowires corresponds to the doping level, i.e., is much lower than the Mn/Ga flux ratio (about 3%) used during the MBE growth. The resistivity measurements of individual nanowires confirmed that they are conductive, in accordance with the photoluminescence measurements which showed the presence of $Mn^{2+}$ acceptors located at Ga sites of the GaAs host lattice of the nanowires. An anomalous temperature dependence of the photoluminescence related to excitons was demonstrated for Mn-doped GaAs nanowires.

**Introduction**

Nanowires (NWs) made from GaAs are attractive for potential applications in nanodevices, since they are built from a material which is widely used in optoelectronics. For functional devices the control over the polarity and concentration of charge carriers, i.e., doping of the given semiconducting materials is essential. However, since the growth of the NWs is governed by principles different from the case of the bulk crystal or thin film growth, the doping mechanisms of a NW are still not fully recognized. Usually the NWs are grown by using a catalyst in the form of liquid nanodroplets (typically gold in the case of GaAs). The formation of NWs involves the dissolution of the constituent elements in the droplet and their accumulation into the growth front of the NW tip, at the droplet/NW interface. In the case of GaAs, the gold nanodroplets widely used to catalyse the NW growth can be replaced by the nanodroplets of Ga, which have the additional advantage of avoiding nonintentional doping of the NWs by Au.[1,2] While the Ga nanodroplets leading to the NW growth cannot be formed on a clean GaAs surface, they can easily be generated if the GaAs surface is covered by a thin $SiO_2$ layer, or on a silicon surface, either oxidized or oxide-free. This method of growing GaAs nanowires driven by Ga droplets was first reported in 2008,[3-5] and since then has been used by various groups, becoming nowadays a standard method for growing self-catalysed GaAs NWs on $SiO_2$ covered GaAs or on Si substrates.[6-10]

In this letter we present the results of investigations of the MBE grown GaAs:Mn NWs. (Ga,Mn)As is the canonical ferromagnetic semiconductor (FMS), with carrier mediated ferromagnetic properties due to diluted substitutions of Ga by $Mn^{2+}$ ions in the GaAs host lattice.[11-14] Mn in this case provides both magnetic moments, due to the spin polarisation of the half-filled 3d shell, and acts as an effective acceptor.[11,12] Typical concentrations of holes in ferromagnetic (Ga,Mn)As are in the range of $10^{20} - 10^{21}$ $cm^{-3}$, for Mn contents above about 1%.

As it was shown previously by us and other groups, the Mn itself can induce the growth of GaAs:Mn nanowires without the presence of any catalyst (Au or Ga).[15-20] However, Mn-catalysed nanowires are very much disordered, tapered, have high tendency for branching and contain a lot of structural defects.[16,19-21] Thus we have exploited the possibility of obtaining the nanowires grown with the other catalyst (Ga droplets in this case) and doped with Mn at high temperature growth, in order to estimate the Mn-doping limits and influence of Mn flux on self-catalysed GaAs NW growth. The $Mn^{2+}$ ions occupying Ga sites in GaAs are shallow acceptors[12] thus Mn can be used as a p-type dopant in GaAs NWs. The p-type doping of GaAs NWs has only recently been demonstrated (p-type doping of GaAs NWs grown by MBE with Be[22] or grown by MOVPE with C[23]). Hence investigations of efficiency of p-type doping of GaAs NWs by Mn are interesting also in this context. The fact that $Mn^{2+}$ ions are also bringing magnetic moments associated with S=5/2 spins of $Mn_{Ga}$ opens possibility of using Mn-doped GaAs NWs for studying interesting phenomena associated with interactions between charge carriers or photons (in case of excitations involving polarized light) with localized, single Mn spins.[24,25] Recent theoretical work by Galicka et. al.[26] has shown that Mn incorporation into GaAs NWs depends on the NW crystallographic structure, hence the zinc-blende phase Mn dopants are preferentially located at the NW side walls and in the hexagonal (wurtzite) phase case they are preferentially located inside the NW.[26] Since it is well known how to grow GaAs NWs in these two crystallographic structures,[3,27] the Mn doping studies in both cases are highly interesting. In this letter we focus on the case of zinc-blende GaAs:Mn NWs; the results for wurtzite GaAs:Mn NWs will be published elsewhere.

## Sample growth and structural properties

### Sample preparation

The GaAs NWs were grown by molecular beam epitaxy on silicon substrates - Si(100) covered with 5 nm thick surface layer of $SiO_2$. Prior to the NW growth the silicon substrate was preheated to about 670 °C for 15 min. Then the substrate temperature was lowered to 630 °C and the GaAs NWs growth was started. As flux was generated by the valved cracker source with the cracking zone temperature of 900 °C, providing arsenic dimers. The $As_2$/Ga flux ratio during the NWs growth was kept close to 0.5, i.e. the NW growth was performed in conditions close to stoichiometric ones. The Mn doping was started shortly after the RHEED patterns typical for NWs were observed typically 10-15 min after the growth starts. The substrate temperature during the Mn doping was lowered from 630 °C to 600 °C, the Mn/Ga flux ratio during doping was close to 3%.

### Structural properties

The NWs were investigated by scanning electron microscopy (SEM) and transmission electron microscopy (TEM). Figure 1 shows SEM images for sample with GaAs:Mn NWs grown at 600 °C.

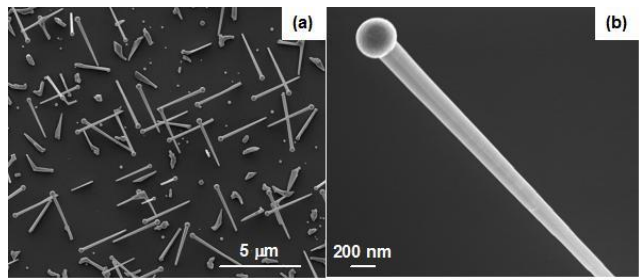

**Figure 1.** Scanning electron microscopy images of GaAs:Mn nanowires grown on Si(100). (a) General (plane) view of a part of the sample surface. The dominant orientation of NWs with surface projections parallel to the Si [011] or [01-1] directions is seen; (b) Magnification view of the upper part of a single NW.

As can be seen in Fig.1a the GaAs:Mn NWs are quite uniform in terms of sizes (diameters and lengths being close to 150 nm and 5 µm, respectively). A slight decrease of the NW diameter with an increasing distance from the NW top can be noticed. The NWs are in epitaxial relation to the Si(100) substrate and grow preferentially along <111> directions. Thus their surface projections are parallel to the orthogonal [0-11] and [011] azimuths of the Si(100) surface. The density of GaAs nanocrystals formed at the substrate surface between the NWs is very low. This is the advantage of the NW growth on oxidized silicon. At high temperatures (600 °C and above) GaAs does not stick to the surface $SiO_2$ layer and the growth takes place predominantly at the regions where the Ga droplets have been formed. Since the NWs were grown at a relatively high Mn/Ga flux ratio of 3%, we have tried to reveal the incorporation of Mn into the NWs. The (Ga,Mn)As solid solution with Mn content above 1%, which is detrimental for ferromagnetic phase transition to occur in this material, can only be grown (by MBE technique) at very low temperatures (in the range of 180 °C - 280 °C). For the growth of NWs, much higher temperatures are required (about 550 °C for Au catalysed growth on GaAs substrates and above 600 °C for self-catalysed growth on Si). The reason for the fact that the layer growth mode of highly Mn doped GaAs at high temperatures is not possible is that above the Mn solubility limit in GaAs the MnAs phase segregation takes place, which inhibits the 2-dimensional layer-by-layer growth.[11,16,20] To our best knowledge, for high temperature grown GaAs NWs, the influence of Mn flux on the NW growth has not been investigated yet.

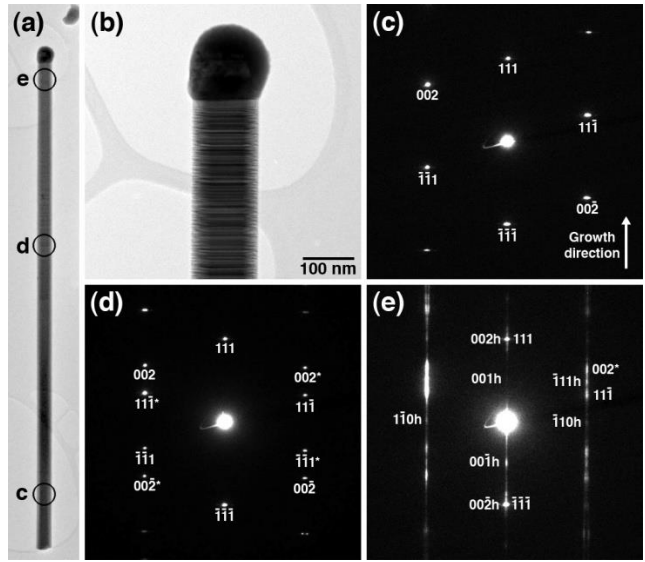

**Figure 2**. (a) and (b) - TEM images and (c) (d) and (e) - electron diffraction patterns collected from the sections of the NW labeled accordingly. The length of the NW shown in (a) is 4.4 μm, the NW has defect-free cubic (zinc-blende) structure in the middle and bottom section, it has a high density of stacking faults and hexagonal (wurtzite) sections in the top part.

Individual nanowires were studied by scanning TEM (STEM) techniques. Fig. 2 shows images of a single NW [the whole NW was removed from Si(100) and placed on a TEM grid], and its top section, together with electron diffraction patterns taken from the NW bottom, middle and top sections labeled (c),

(d) and (e), respectively. The NW has predominantly zinc-blende structure. The bottom and middle sections are defect-free, whereas the top region close to the Ga droplet has very high density of stacking fault defects and, accordingly short hexagonal (wurtzite structure) sections.

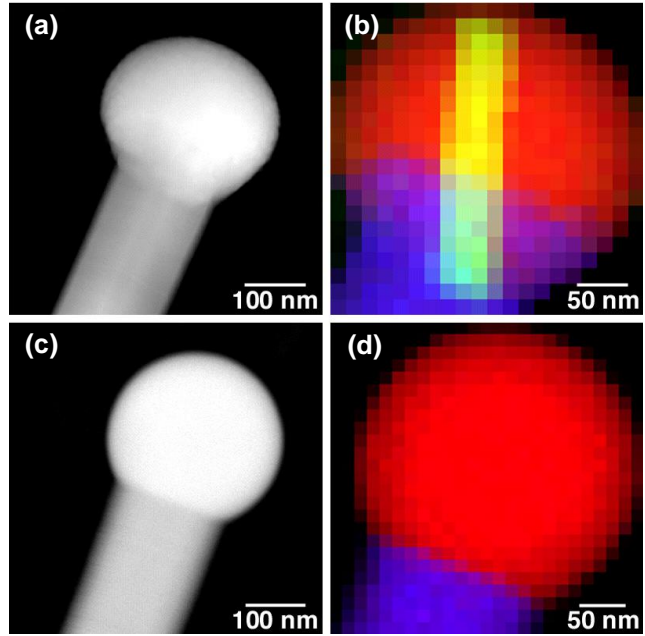

**Figure 3.** (a) HAADF TEM image of the top part of GaAs:Mn nanowire and (b) EDS chemical map of the Ga droplet and interface region between the Ga droplet and the nanowire. Different elements are represented by the different colors: Ga – red, As - blue, Mn - green. Yellow stripe inside the Ga droplet is the Mn-rich region; the stripe color is due to the overlap of red and green representing Ga and Mn, respectively. In comparison the same analysis has been performed for undoped GaAs NW - (c) and (d), and no Mn-rich region in the Ga droplet has been found.

Figure 3 shows a high-angle annular dark-field (HAADF) TEM image of the Mn-doped NW [(a)], and elemental maps acquired with energy-dispersive spectroscopy (EDS) of its top part and the Ga droplet [(b)]. The presence of the Mn-rich region inside the Ga droplet and at the droplet-NW interface region is well seen. For comparison the pure GaAs NWs were characterized in the same manner (Fig. 3c and 3d) and no Mn inside the droplet was found in this case. The signals from the elements of As, Ga and Mn in the EDS maps are coded in blue, red and (false) green colors. Yellow regions (overlap of red and green) due to the presence of Mn and Ga are clearly visible in NWs doped with Mn. As can be seen in Fig 3b Mn is detected inside the Ga droplet only for GaAs:Mn NW.

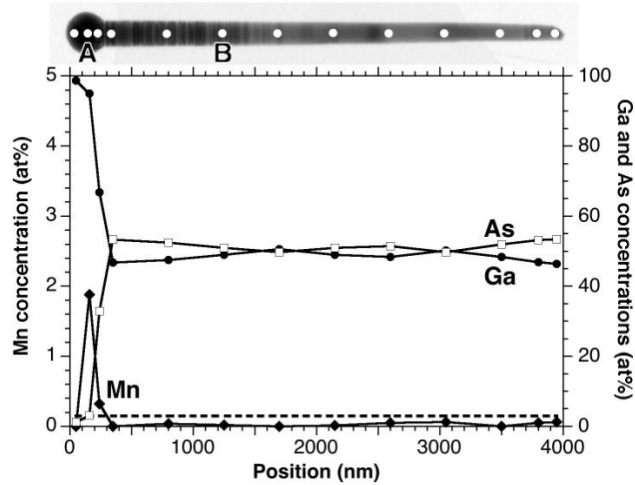

**Figure 4.** Semi-quantitative EDS point analysis along the GaAs:Mn NW shown in the upper panel. The unit in the vertical axis is at%. The dotted line shows the detection limit of EDS for Mn.

The EDS line scan along the top part of the NW shown in Fig.4 confirms the high Mn percentage inside the Ga droplet and demonstrates that in the NW body the Mn content is below the detection limit of EDS method (i.e. below 0.15 at %). As it is known, several solid solution structures corresponding to various Mn to Ga ratio are known for the Mn-Ga system. Unfortunately, present data do not allow identifying the exact solid phase of Mn-rich region in the Ga droplet and this point requires further, more detailed studies.

**Optical properties**

The search for Mn using the TEM methods did not demonstrate the presence of Mn in the NWs.[21] Thus we have performed optical measurements (PL spectroscopy, CL spectroscopy, and micro-Raman scattering) in search for Mn-related structures or at least for possible Mn-related influence on the optical spectra.

**Photoluminescence**

In GaAs, Mn doping gives rise to a well-known mixing of a donor-acceptor pair (DAP) transition (A being the manganese acceptor) and a conduction band-Mn acceptor (e, Mn) transition.[28] This transition is usually observed in PL spectra as a characteristic zero-phonon emission line at ~ 1.41 eV (activation energy of Mn acceptor $E_a \approx 110$ meV).[24,28-33] The detection limit of the luminescence connected with Mn acceptor in GaAs is of the order of $10^{13}$ cm$^{-3}$.[34] The PL spectra were measured using a Sharmrock SR-303i spectrometer with the spectral resolution close to 0.2 nm. The PL signal from the sample was accumulated by CCD detector. The samples were mounted on a cold finger of a closed-cycle helium cryostat and excited by the 532 nm and 325 nm lines from Nd:YAG and He-Cd lasers, respectively. The PL measurements were performed at various temperatures from 5 K to 200 K in a macro mode. Thus the detected signal originated from hundreds of NWs at one time (and also from possible GaAs nanocrystals at the Si substrate between the NWs). The excitation power dependent measurements were performed at the laser power of 50 μW – 20 mW. The excitation laser power was varied by means of neutral density filters. The spectral positions of the PL peaks were fitted by using a multiple Lorentzian line shapes. Although the collected spectra from different locations on the sample, and thus from bunches of different NW, were generally similar, we do see variations in relative intensities and positions (up to 3 meV) of the particular peaks. We attribute this variability to the variation of the NW morphology and thus to the number and energetic position of the surface states at the NW edges or different density of defects.

In Fig. 5 (a) we compare the PL spectra measured at 5 K for GaAs and GaAs:Mn NWs. These spectra were excited with the 532 nm laser line and at excitation intensity of 20 mW. The best pronounced, common feature of both spectra is an emission band located in 1.45 eV – 1.5 eV spectral range which consists of two peaks. The high energy peak is located at 1.485 eV for both samples whereas the low energy peak is at 1.473 eV for GaAs NWs and is (slightly) shifted to 1.463 eV for GaAs:Mn NWs. Additionally for Mn-doped NWs a new peak appears at the energy of 1.410 eV. This peak is accompanied by two additional bands at lower energies, which can be attributed to its LO-phonon replicas. The emission lines which occur at 1.485 eV, 1.463 eV and 1.410 eV shift to lower energy with decrease of excitation intensity, which is characteristic for donor-to-acceptor pair (DAP) transitions, what is evidenced in Fig. 6. In contrary the energy of the emission line due to exciton bound to neutral donor ($D^0$, X)[35] observed at about 1.514 eV for both samples, is almost power independent. In particular, the emission band centered near 1.48 eV is commonly attributed to free-to-bound (FB) and DAP transitions involving Si acceptors.[36-41] Skromme et al.[36] observed at 2 K an emission line at 1.4816 eV in the PL spectrum measured for the vacuum phase epitaxy-grown GaAs and interpreted it as a $Si_{As}$-associated DAP transition. The luminescence centered at 1.483 eV was also observed in the low-temperature PL spectra for a MBE-grown Si-doped GaAs[37,38] and it was also attributed to the Si-donor to Si-acceptor transition. Thus the PL emission centered at 1.485 eV for investigated NWs is more likely a feature associated with the Si acceptor. The presence of Si impurities in NWs grown with Ga droplets playing the role of the catalyst is highly possible due to the NWs growth procedure used. It has been shown[42,43] that before the NW growth is initiated, the Ga droplets are contacted with the Si through the pinholes in the $SiO_2$ surface layer. Since, prior to the NW growth, the Ga droplets are formed at the hot Si surface the diffusion of Si into the droplet is highly probable. The Ga droplets are even used as catalyst for growth of Si NWs[44] at significantly lower temperatures then used by us to grow GaAs NWs described here. The discrepancies between positions of the transitions lines observed in the literature are due to different experimental conditions. DAP peak positions generally strongly depend on excitation power density and on dopant concentration, for higher doping levels the FB and DAP peaks can also overlap. On the other hand, carbon, the most frequently introduced residual contaminant in every growth technique acts as an acceptor in GaAs with a characteristic transition in the PL spectrum located at slightly higher energies (about 1.49 eV) than that of Si.[36] The observed small shifts of transition frequencies due to the presence of local electric fields generated on the NWs surface by charged states make their unambiguous identification difficult so we cannot definitely rule out a certain contribution to our PL spectra resulting from possible C content in investigated samples. The origin of the luminescence centered at about 1.46 eV remains unclear. Fang et. al.[40] and Das et. al.[41] attributed this emission to some non-identified additional impurity, whereas Parchinskiy et. al.[45] suggested a Ga antisite related transition. The PL peaks at 1.467 eV and 1.453 eV were observed by Wagenhuber et. al.[30] in MBE GaAs/Si and have not been identified. Nevertheless, our power dependent PL measurement indicates that this line is due to DAP transitions.

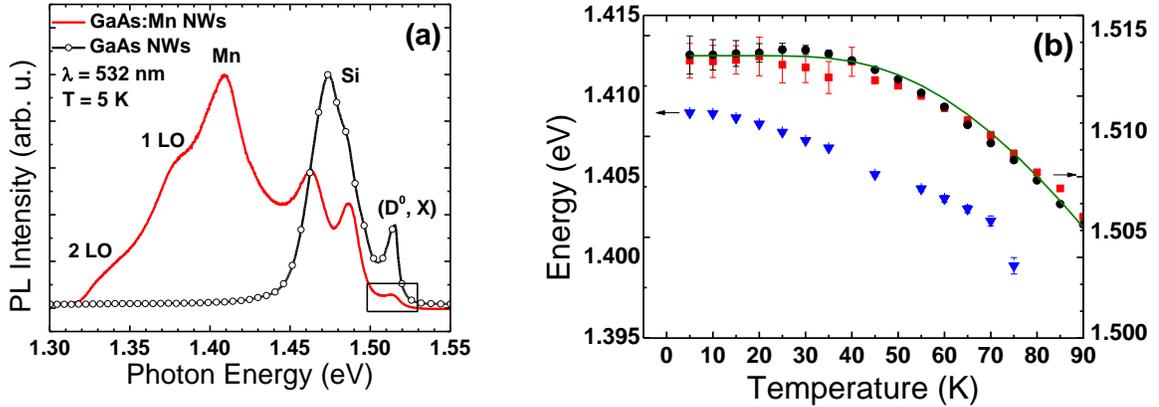

**Figure 5.** (a) PL spectra (normalized to the highest intensity) for GaAs and GaAs:Mn NWs excited by the 532 nm laser line. The part of the spectrum marked in the Fig. 5 (a) by the rectangle is presented in Fig. 8 and analyzed further in the text. (b) Temperature dependence of the excitonic transition energy (red squares – GaAs:Mn NWs, black circles – GaAs NWs) and Mn-related transition in GaAs:Mn NWs (blue triangles) plotted together with semi-empirical Bose-Einstein model (solid line).[46]

($D^0$, X) emission line is much more pronounced in the case of GaAs NWs. It is possibly due to the enhanced compensation of the residual donors by activated Mn acceptors. The temperature dependence of these transitions observed for the NWs investigated here (see Fig. 5(b)) can be well described by the semi-empirical Bose-Einstein model proposed by Vina et al.[46]

$$E(T) = E_B - a_B[1+2/(\exp(\theta/T) - 1)] \qquad (1)$$

where $a_B$ represents the strength of the electron-phonon interaction and $\theta$ corresponds to the average phonon frequency

The temperature dependence of Mn related transitions energy is similar to that corresponding to the exciton line.

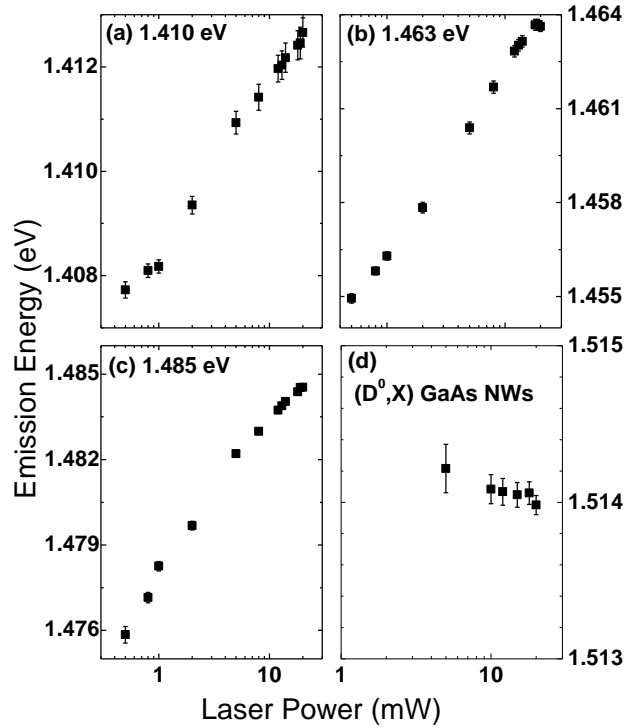

**Figure 6**. Laser power dependence of impurity related PL peaks: (a) 1.410 eV, (b) 1.463 eV and (c) 1.485 eV for GaAs:Mn NWs, and (d) of excitonic emission ($D^0$,X) from GaAs NWs

The comparison of relevant PL spectra excited with the 325 nm laser line similar to those shown in Fig. 5a is shown in Fig. 7. Also for this excitation energy for Mn-doped NWs a new peak appears at the energy of about 1.41 eV. This is a well-known peak observed previously in (Ga,Mn)As[24,30-33] and can be interpreted as a recombination via an acceptor level of the Mn impurity located in the Ga sublattice. Although Stolz et al.,[47] observed the 1.411 eV emission line in the PL spectra of undoped and Si-doped GaAs layer MBE-grown on GaAs and Si substrate and attributed it as a defect-to-carbon acceptor recombination process, the emission band near 1.40 eV is not present for the GaAs NWs, which leads us to the conclusion that this PL peak is associated with Mn acceptor.

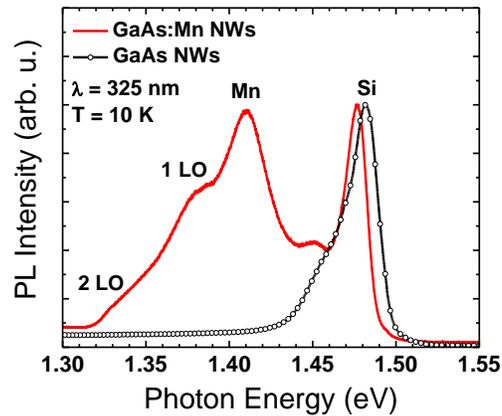

**Figure 7**. PL spectra for GaAs and GaAs:Mn NWs excited by a 325nm laser line. PL intensity is normalized to that of the 1.48 eV peak

It is worth to notice that the intensity of the Si related PL at about 1.48 eV significantly increases as compared to the intensity of the peak around 1.46 eV when excited by the 325 nm laser line. Possible explanation of observed effect is related to the high absorption coefficient value in the UV region for GaAs which limits the PL excitation to the narrow NWs slab close to the surface. This is that part of the NW where an accumulation of Si dopants has been theoretically predicted in zinc blende GaAs NWs.[48] The observed small energy shift in the Si PL (about 5 meV) seen here is due to an influence of a local electric field on the Si level position resulting from the presence of the surface states. An interesting phenomenon is observed when studying the temperature dependence of the exciton PL intensity. For GaAs NWs this intensity decreases in a typical manner with an increasing temperature. However, in the case of GaAs:Mn NWs quite unexpected temperature dependence of the 1.514 nm PL line intensity is found. The comparison of PL spectra taken for GaAs:Mn NWs at few different temperatures is presented in Fig.8 (only a high-energy part of each spectrum is shown).

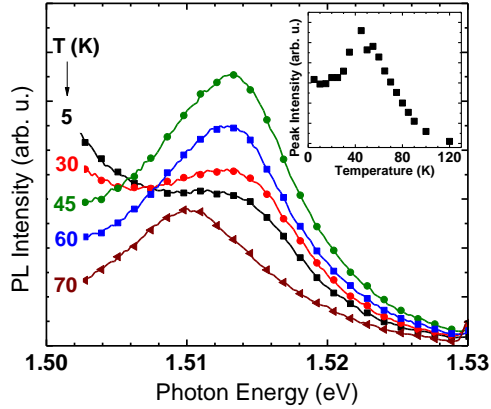

**Figure 8**. Temperature evolution of the exciton line observed for GaAs:Mn NWs (laser excitation line 532 nm, excitation power 20 mW). The inset shows variation of the intensity of the peak shown in the main frame, in the broader temperature range.

An anomalous temperature dependence of the exciton emission intensity can be explained as follows. Due to the amphoteric behavior of Si dopant in GaAs NWs both $Si_{Ga}$ donors and $Si_{As}$ acceptors can be created during the NWs growth.[49] In the case of low impurity concentration the excitons creation is very probable and the temperature activation of a small number of acceptors (expected for GaAs NWs) has no influence on the total intensity of the exciton emission. The situation changes when a significant concentration of Mn acceptors is additionally present in investigated NWs. At the lowest temperatures both the compensation of Si donors by acceptors and relatively high Mn acceptor ionization energy (110 meV) strongly reduce a free-carrier concentration and a probability of exciton creation. Because of that the intensity of PL exciton line observed at 5 K is very low. This intensity increases with temperature due to the activation of acceptors resulting in an increase the of free-hole concentration. Starting from some temperature free holes resulting from acceptor ionization have no more influence on the exciton population and one come back to the standard situation (PL intensity decrease). The result presented in Fig. 7 is an additional evidence of the presence of Mn acceptors in investigated GaAs:Mn NWs. It can be pointed out that a trace of similar anomalous temperature dependence of exciton emission has been reported previously (without an explanation) for GaAs:Si.[35] However, in that case this effect was much less pronounced and concealed by an intense exciton emission. Because of the relatively large diameters of the NWs we do not expect any confinement effects. Both the length and the diameter of the NWs considerably exceed the relevant length scales; for example the exciton Bohr radius in GaAs is in the range of 9-13 nm.[50] We have also not detected any features in the PL spectra which could be attributed to the regions of NWs with the wurtzite structure.[51]

**Cathodoluminescence**

In parallel to PL the same samples were also investigated by cathodoluminescence (CL) at 4 K (see Fig. 9).

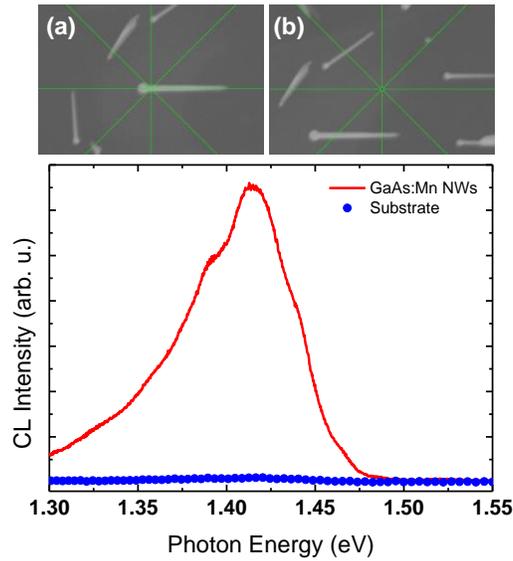

**Figure 9.** (a) and (b) – SEM images of the GaAs:Mn NWs. The scale bars on the SEM pictures correspond to 1 μm. The middle of the crosshair indicates the position from which the spectra on panel (c) have been collected. (c) CL spectra of GaAs:Mn NW – solid (red) line, and of the substrate between the NWs – circled (blue) curve. The excitation energy 10kV, T = 4 K.

The exact shape of the PL lines cannot be reproduced by the CL spectra because of a huge charge accumulation in investigated NWs during the CL measurements and to some extent due to different area of NWs excited using both experimental techniques. Nevertheless, the principal information resulting from both methods was the same. The micro-CL measurements performed separately on the single NWs and on the sample surface between them (seen at the SEM pictures in Fig. 9 (a) and (b), respectively) proved definitely that the PL signal came from the NWs only.

**Raman scattering**

Additionally we have performed Raman scattering measurements for the investigated NWs. All Raman scattering measurements were performed in a quasi-backscattering geometry using a Jobin-Yvon U1000 spectrometer equipped with a microscope, holographic gratings, a S20 photomultiplier, and a photon counting system. The Ar$^+$ laser lines with the wavelength of 488 nm and 514.5 nm were applied for the excitation. The Raman spectra corresponding to as-grown GaAs:Mn NWs on the Si substrate were collected at room temperature with the spectral resolution close to 1 cm$^{-1}$. Analogous spectra were also taken for pure GaAs NWs for comparison. Depending on the microscope objective (x100 or x50) a single NW or a group of two or three NWs were illuminated and investigated by the micro-Raman scattering method.

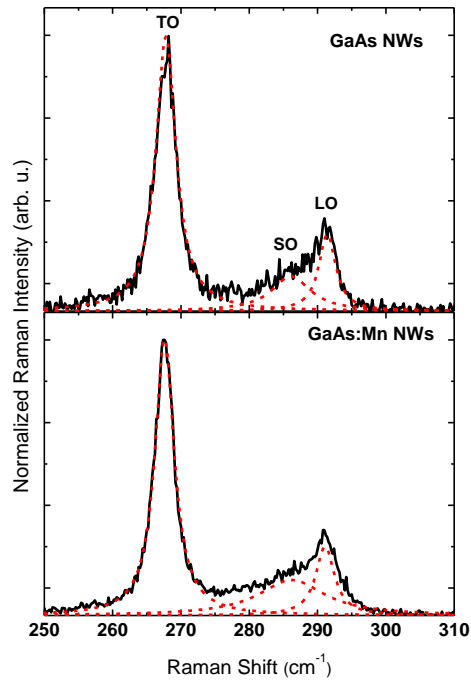

**Figure 10.** Room-temperature Raman spectra taken for a group of two GaAs:Mn NWs and for a single GaAs NW (514.5 nm laser line, x50 microscope objective). Thick solid curve shows the experimental data. The assumed three independent phonon contributions (TO, SO, and LO), are represented by dashed lines.

Figure 10 shows a typical example of a room temperature micro-Raman spectrum taken on a group of two GaAs:Mn NWs side by side. Two pronounced, well separated structures in this spectrum correspond to TO and LO phonon excitations, respectively. Just below the LO phonon frequency a broad structure with a small intensity can be seen. This structure can be attributed to the presence of a surface optical (SO) phonon mode. Such a structure has been reported previously for GaAs NWs.[6] The spectra taken on one single GaAs:Mn NWs were almost identical except an obviously lower signal intensity and a resulting poorer signal to noise ratio. The observed spectrum is quite similar to that of a single pure GaAs NW, taken for comparison. The values of both, frequencies and FWHM's of particular phonon excitations observed for GaAs:Mn NWs do not differ significantly from those observed for GaAs NWs. Thus, the Mn concentration in GaAs:Mn NWs should not exceed 0.05% ($1\cdot10^{19}$ cm$^{-3}$).[52] The only tiny difference in the Raman spectra between pure and the Mn-doped GaAs NWs is slightly higher FWHM for the SO mode in the Mn-doped GaAs NWs as compared to that of pure GaAs NWs (11.8 cm$^{-1}$ and 6.9 cm$^{-1}$, respectively). Some details of Raman spectrum shown in Fig. 10 (in particular, the SO mode frequency and its FWHM value) cannot be fully explained assuming a pure zinc-blende structure of the investigated NWs. The simultaneous presence of several segments with the wurtzite structure and with the zinc blende structure in the same NW should result in a small frequency shift of all phonon modes: TO, SO, and LO, as it was recently demonstrated.[53] A presence of wurtzite segments in GaAs:Mn NWs under study was shown by TEM (Fig. 2). However it is hard to to estimate the content of segments with the wurtzite structure in the total NW volume contributing to the Raman spectra, from the TEM images acquired for a specific single NW. Nevertheless, the area with the high concentration of structure defects and stacking faults is well seen in the upper part of all investigated NWs. Not only the concentration of structure defects but also the diameter noticeably changes along the NW. The non-uniform distribution of the wurtzite segments and structure defects as well as modifications of the diameter along the NW could explain the observed form of structure below LO phonon line and its anomalously high FWHM value. Taking into account the fact that a whole NW (or NWs) volume is illuminated by the laser light the hypothesis that this structure is due to a superposition of several slightly shifted SO phonon modes corresponding to different NW area is very probable.

**Transport properties**

Since the Mn located at Ga sites in GaAs host lattice is an acceptor,[11,12,54] the Mn-doped GaAs NWs are expected to be conductive, hence the transport properties of single NWs have been investigated. Individual NWs were picked up from the as grown sample and contacted with 4 ohmic contact stripes fabricated by standard nano-litographical techniques (see Ref. 55 for details). The NWs placed on pre-patterned chips are cleaned from oxide by etching in diluted HCl acid at 25 °C for 15 seconds. In addition to this, passivation of the NWs surface was done by a wet etch of ammonium sulfide (NH$_4$)S$_x$ at 25 °C for 2 minutes. After the wet etch the chip with the NWs was transferred to the Physical Vapor Deposition (PVD) system for metallization. Metal contacts consist of Pd/Zn/Pd layers which provide good contact to p-type GaAs NWs.[56]

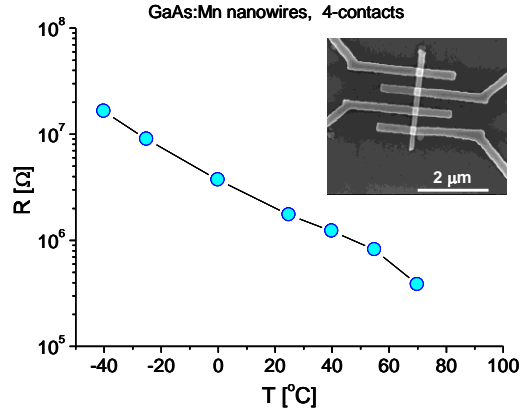

**Figure 11.** Temperature dependence of resistivity measured in a four-contact configuration for a single GaAs:Mn NW, at temperatures around 0 °C. The SEM picture of the NW device with four contacts used for the measurements is shown in the inset.

Figure 11 shows a typical temperature dependence of resistivity for such a NW, around 0 °C. In contrast to the undoped GaAs NWs which are highly resistive (with a resistivity impossible to measure using conventional techniques), the resistivities in the range of 0.1 – 10 MΩ have been typically observed in GaAs:Mn NWs by us (for the same temperature range).

The dependence of the I-V characteristics on the voltage applied to the substrate back gate was also investigated and confirmed the p-type conductivity, corroborating the acceptor character of Mn embedded into NW.[55] The room temperature resistivities in the range of 1 MΩ are comparable to those observed in the Mn implanted GaAs NWs of similar (slightly lower) diameters, grown by MOVPE with Au catalyst, reported recently.[57,58] However neither in NWs obtained via Mn doping of GaAs NWs at high-temperature MBE growth (this paper) nor in Mn implanted and low-temperature annealed GaAs NWs presented in Ref. 57 and 58, a diluted (Ga,Mn)As alloy with a Mn content high enough to allow the emergence of a FM phase transition was obtained.

In our results the Mn doping limit during high-temperature NW growth is much below the occurrence of a FM phase transition in (Ga,Mn)As (corresponding to about 1% of Mn in cation sublattice, i.e. in the range of $10^{20}$ cm$^{-3}$). Even though the FM phase transition, with critical temperature ($T_c$) in the range of 40 -100 K has been observed in Mn-ion implanted and pulsed laser annealed GaAs layers,[59,60] the same high quality FM (Ga,Mn)As solid solution has not yet been obtained in case of Mn ion-implanted NWs;[57,58] i.e. neither the FM phase transition, nor resistivities comparable to those typical for high quality (Ga,Mn)As layers have been demonstrated. This may be due to the very narrow window for thermal treatment (post-implantation annealing) necessary to perform after Mn-ion implantation to reduce the implantation-induced defects. It is well known that the post-growth annealing needed to out-diffuse the Mn from interstitial sites in (Ga,Mn)As and increase its conductivity and FM phase transition temperature is a very delicate process, and is typically performed at temperatures as low as 140 – 180 $^{o}$C.[61-63] 250 $^{o}$C post-implantation annealing used by Paschoal et. al.[58] could actually cause degradation of (Ga,Mn)As due to the possible diffusion of Mn from Ga sites and/or Mn clustering. Recently it has been shown that 400 $^{o}$C annealing leads to complete decomposition of (Ga,Mn)As ternary alloy,[64,65] and that this degradation has already started at much lower annealing temperatures (in the range of 250 – 300 $^{o}$C).[66] Thus it seems that, to day, the only way to obtain (Ga,Mn)As in NW geometries is the growing of GaAs-(Ga,Mn)As core shell structures[21,67] with GaAs core NWs grown at optimum conditions and (Ga,Mn)As shells grown at conditions optimized for (Ga,Mn)As layers, which is are well known after almost two decades of extensive research devoted to this ferromagnetic semiconductor.[68]

**Conclusions**

In conclusions, investigations of GaAs nanowires grown on oxidized Si(100) substrates, doped with Mn at high temperature MBE growth have shown that the Mn doping does not affect the NW growth, even though the Mn to Ga flux ratio during the NW growth is as high as 3%. It has been found by photoluminescence and cathodoluminescence spectroscopy, as well as by transmission electron microscopy, that only a small fraction of Mn is incorporated into GaAs NWs. The Mn concentration along the NW being probably of the order of $10^{18}$ cm$^{-3}$ is much lower than that expected from the Mn/Ga flux ratio set during the MBE growth. This finding is in agreement with the micro-Raman scattering spectra, taken for the single NW. The temperature dependence of the resistivity also confirms the semiconductor characteristics of the electron transport in such NWs with the Mn concentration in a doping level. TEM analysis of the composition of NW in close to the droplet region has shown that excess Mn is accumulated inside the Ga droplets. Photoluminescence measurements revealed the presence of transitions related to the Si defects both in undoped and Mn-doped NWs. In the latter case transitions associated with Mn acceptor states were also detected. An anomalous temperature dependence of the exciton emission was found for GaAs:Mn NWs and explained by the thermal activation of Mn acceptor states.

**Acknowledgements**

The MBE system used for the growth of NWs has been supported by the Swedish Research Council (VR). KG gratefully acknowledges financial support from the European Union within European Social Fund through Human Capital Programme. JS acknowledges partial support by the FunDMS Advanced Grant of the ERC. KG, TW, WZ and WS acknowledge partial financial support by the European Union within European Regional Development Fund, through grant Innovative Economy (POIG.01.01.02-00-008/08). AS acknowledges the financial support by the European Commission Network SemiSpinNet (PITN-GA-2008-215368) and by the European Regional Development Fund through the Innovative Economy grant (POIG.01.01.02-00-108/09). HQX acknowledges the supports from the Swedish Research Council (VR) and the National Basic Research Program of the Ministry of Science and Technology of China (Nos. 2012CB932703 and 2012CB932700).

**Corresponding author**

Janusz Sadowski

Email address: janusz.sadowski@maxlab.lu.se

‡ J.F. Morhange is now retired; he can be contacted at jf.morhange@orange.fr.